\documentstyle[12pt]{article}
\normalbaselines

\newcommand{\be}{\begin{equation}}
\newcommand{\ee}{\end{equation}}
\newcommand{\ba}{\begin{eqnarray}}
\newcommand{\ea}{\end{eqnarray}}
\newcommand{\baa}{\begin{eqnarray*}}
\newcommand{\eaa}{\end{eqnarray*}}
\newcommand{\bb}{\begin{thebibliography}}
\newcommand{\eb}{\end{thebibliography}}

\newcommand{\lab}[1]{\label{#1}}
\newcommand{\re}[1]{(\ref{#1})}

\begin{document}

\begin{flushright}
  UdeM-LPN-TH-113/92
\end{flushright}

\vbox {\vspace {4mm}}
\begin{center}

{\Large \bf Exactly Solvable Potentials and Quantum Algebras}%
\footnote{Talk presented at the XIXth International Colloquium on
Group Theoretical Methods in Physics, Salamanca (Spain),
June 29 - July 4, 1992}\\[5mm]
V.Spiridonov%
\footnote{On leave of absence from the Institute for Nuclear Research,
Moscow, Russia}\\[1mm]
{\it Laboratoire de Physique Nucl\' eaire,
Universit\' e de Montr\' eal, \\
C.P. 6128, succ. A, Montr\' eal, Qu\' ebec, H3C 3J7, Canada}\\[5mm]
\end{center}

\begin{abstract}
Self-similar potentials and corresponding symmetry algebras are
briefly discussed.
\end{abstract}

The following scheme describes applications of quantum-algebraic
structures within the context of ordinary quantum mechanical spectral
problems.  Let $T$ be an operator of the affine transformation on line
\be
T f(x) = \sqrt{q} f(qx+a), \qquad x\in R,
\lab{e1}
\ee
where $0<q\leq 1$ and $a\in R$ are scaling and translation parameters
respectively. $T$ is unitary in $L^2$-space, $T^\dagger=T^{-1}$.
We introduce two factorization operators
\be
A^+=(p+iW(x))\,T, \qquad
A^-=T^{-1} (p-iW(x))=(A^+)^\dagger,
\lab{e2}
\ee
where $W(x)$ is an arbitrary real function (superpotential).
They define two Hamiltonians
$$
H_-=A^+A^-=p^2+W^2(x)-W^\prime(x)=p^2+U_-(x),
$$
\be
H_+={1\over q^2}A^-A^+=p^2+{1\over q^2}W^2({\mbox{$\frac{x-a}{q}$}}) +
{1\over q}W^\prime({\mbox{$\frac{x-a}{q}$}})=p^2+U_+(x).
\lab{e3}
\ee
These may be unified into one $2\times 2$ Hamiltonian
\be
{\cal H}=\left(\matrix{H_-&0\cr 0&H_+\cr}\right)=
p^2+V(x)+B(x)\sigma_3,\qquad \sigma_3=\left(\matrix{1&0\cr 0&-1\cr}\right),
\lab{e4}
\ee
where $V(x)$ and $B(x)$ are linear combinations of $U_\pm(x)$.
Introducing supercharges
$$
{\cal Q}^-=\left(\matrix{0&0\cr A^-&0\cr}\right), \qquad
{\cal Q}^+=\left(\matrix{0&A^+\cr 0&0\cr}\right),
$$
we obtain a deformation of the $sl(1|1)$ superalgebra
\be
{\cal Q}^+{\cal Q}^- +q^{-2}{\cal Q}^-{\cal Q}^+={\cal H}, \quad
({\cal Q}^\pm)^2=0,\quad
{\cal H}{\cal Q}^\pm=q^{\pm 2} {\cal Q}^\pm {\cal H}.
\lab{e5}
\ee
The limit $q\to 1$ restores conventional relations. However, even in this case
we have a generalization of the standard supersymmetric models because
the shift parameter $a$ enters the potential $U_+(x)$,
i.e. still $A^\pm$ are infinite order differential operators.

Comparing \re{e5} with the undeformed algebra one can see that double
degeneracy is lifted, the energy split is proportional to $1-q^2$.
Subhamiltonians $H_\pm$ are not isospectral but their eigenvalues
are related via scaling. Vacuum energy of $\cal H$ is formally semipositive,
$E_0\geq 0$, the equality is reached when $A^-$ (or $A^+$) has
a normalizable zero mode.

Let us take ``magnetic field" $B(x)$ in \re{e4} to be homogeneous,
$U_-(x)-U_+(x)=2B=const.$ This leads to the mixed
finite-difference-differential
equation determining superpotential,
\be
W^\prime(x)+qW^\prime (qx+a)+W^2(x)-q^2 W^2(qx+a)+2Bq^2=0.
\lab{e6}
\ee
For convenience we rewrite \re{e6} in the form
\be
f^\prime(x)+qf^\prime (qx)+f^2(x)-q^2 f^2(qx)=k,
\lab{e7}
\ee
where $f(x)=W(x+a/(1-q))$ and $k=-2Bq^2$. Parameter $k$ may be removed
from \re{e7} by rescaling  but we keep it as a unique (positive) dimensional
parameter. Note also that if $x$ and $q$ are complex variables
then it is natural to consider only  $|q|\leq 1$ region due to the
symmetry $f(x,q^{-1})=$ $iqf(qx/i,q)$.

For this special case, one can write instead of \re{e3}
\be
A^+A^-=H+k/(1-q^2), \qquad A^-A^+=q^2H+k/(1-q^2),
\lab{e8}
\ee
where
\be
H=p^2+f^2(x)-f^\prime(x)-k/(1-q^2).
\lab{e9}
\ee
Now it is easy to see that $A^\pm$ and $k$ form a $q$-deformed Heisenberg-Weyl
algebra [1],
\be
A^-A^+-q^2A^+A^-=k, \qquad [A^\pm, k]=0.
\lab{e10}
\ee
The Hamiltonian $H$ $q$-commutes with $A^\pm$
\be
H A^\pm=q^{\pm2} A^\pm H.
\lab{e11}
\ee
As a result, from the normalized vacuum state $|0\rangle$, $A^-|0\rangle =0$,
a series of energy eigenstates is generated
$$
|n\rangle={(A^+)^n\over \sqrt{k^n[n]!}} |0\rangle, \qquad
\langle n|m\rangle=\delta_{nm}, \quad
[n]!=\prod_{i=1}^n{1-q^{2i}\over 1-q^2},
$$
$$
A^+|n\rangle =k^{1/2}\sqrt{\,{1-q^{2(n+1)}\over 1-q^2}}\,
|n+1\rangle ,\qquad
A^-\, |n\rangle =k^{1/2}\sqrt{\, {1-q^{2n}\over 1-q^2}}\,
|n-1\rangle,
$$
\be
H\,|n\rangle=E_n|n\rangle,\qquad E_n=- {k\over 1-q^2} q^{2n}.
\lab{e12}
\ee
This gives the whole discrete spectrum provided $A^\pm$
respect boundary conditions of the problem.

It is convenient to consider antisymmetric superpotential,
$f(-x)=-f(x)$, which is given by the power series
\be
f(x)=\sum_{j=1}^{\infty} c_j\, x^{2j-1}, \qquad
c_j={q^{2j}-1\over q^{2j}+1}{1\over 2j-1}\sum_{l=1}^{j-1}c_{j-l}c_l, \quad
c_1= {k\over 1+q^2}.
\lab{e13}
\ee
Odd and even $c_j$ have different signs and the series converges
for arbitrarily large $x$ with $f(\infty)>0$. After analytical continuation
of \re{e13} to the imaginary axis (or, for $q>1$) all expansion coefficients
are positive and a pole singularity develops at some point $x_0$. In both
cases the state $|0\rangle\propto \exp (-\int^xf(y)dy)$ is normalizable by
topological arguments (for the singular superpotential the space region
should be restricted to $[-x_0,x_0]$). Only first solution corresponds to
\re{e12}. In the second case operator $T$ moves positions of poles and zeros
so that $A^+$-action brings singularities into wave functions. Therefore
\re{e12} is not valid for $q>1$.

Various limits of $q$ recover well-known exactly solvable potenials.
At $q=0$ a subcase of the Rosen-Morse problem arises, $f(x)\propto \tanh x$.
Analytical continuation (or, the limit $q\to \infty$ at fixed $c_1$) gives
P\"oschl-Teller problem, $f(x)\propto \tan x$. The limit $q\to 1$ restores
customary harmonic oscillator potential (note that for $W(x)$ this is a
non-trivial limit). If $q$ is complex, then hermiticity properties of
$A^\pm$ are broken. Nevertheless, there are interesting physically relevant
real potentials at $q^n=1$. The $q=-1$ case is equivavlent to $q=1$.
At $q^3=1$ the Schr\"odinger equation coincides with the simplest Lam\'e
equation for equianharmonic Weierstrass function. Corresponding spectral
problem, defined by the requirement for wave functions to vanish in singular
points, is known to be solvable. This example provides ordinary differential
calculus realization of cyclic representations of the $q$-oscillator
algebra \re{e10}, the associated $q$-special functions being the ordinary
Lam\'e functions.

Equation \re{e7} was found by A.Shabat [2] after substitution of the
self-similarity ansatz
$f_j(x)=q^jf(q^jx),\; k_{j}=q^{2j}k,$ into the chain of coupled
Ricatti equations
\be
f_j^\prime(x)+f_{j+1}^\prime(x)+f_j^2(x)-f_{j+1}^2(x)=k_j,\qquad j\in Z,
\lab{e14}
\ee
which arises in factorization method. He also described general structure
of this system for real $x$ and $0\leq q <1$ through the $n$-soliton,
$n\to \infty$, approximations. Connection with the $q$-oscillator algebra was
established in Ref.[3]. Associated deformation of supersymmetric
quantum mechanics was discussed in Ref.[4]. Appearance of the
equianharmonic Weierstrass function at $q^3=1$ was described in Ref.[5].

In the rest part of this paper we outline a generalization of the
Shabat's potential. First we note that periodic conditions
\be
f_{j+N}(x)=f_j(x), \qquad k_{j+N}=k_j,
\lab{e15}
\ee
provide finite-dimensional truncation of the chain \re{e14}. Odd $N$ cases,
endowed by the restriction $\sigma\equiv k_1+k_2+\dots +k_N=0,$ are known
to be integrable in terms of the hyperelliptic functions defining
finite-gap potentials [6]. If $\sigma\neq0$, then one has essentially more
complicated situations. At $N=1$ this gives harmonic oscillator problem.
The $N=2$ system coincides with the conformal quantum mechanical model,
\be
f_{1,2}(x)={1\over 2}(\pm\, {k_1-k_2\over k_1+k_2}{1\over x} +
{k_1+k_2\over 2}x).
\lab{e16}
\ee
Already $N=3$ case leads to transcendental potentials, namely, $f_j$ start
to depend on the solutions of Painlev\'e-IV equation [7].

Applying the ideas of deformed supersymmetry to the whole chain \re{e14}
the author have found the following $q$-periodic closure
\be
f_{j+N}(x)=qf_j(qx), \qquad k_{j+N}=q^2 k_j.
\lab{e17}
\ee
These conditions describe $q$-deformation of the finite-gap and related
potentials discussed in Refs.[6,7]. The Shabat's system is generated
at $N=1$. Because of the highly transcendental character of self-similarity
and relation with the Painlev\'e equations, all potentials associated to
\re{e17} may be called as $q$-transcendental ones.

Algebraically, $q$-transcendental potentials are characterized by the
ladder relations
\be
HA^+-q^2A^+H=\sigma A^+,\qquad A^-H-q^2 H A^-=\sigma A^-,
\lab{e18}
\ee
$$A^+=(p+if_1)(p+if_2)\dots(p+if_N)T,\qquad A^-=(A^+)^\dagger, $$
$$H=(p+if_1(x))(p-if_1(x)),$$
and the identities
\be
A^+A^-=\prod_{i=1}^N (H-\lambda_i), \qquad
A^-A^+=\prod_{i=1}^N (q^2 H+\sigma -\lambda_i),
\lab{e19}
\ee
where parameters $\lambda_i$ are defined by the equalities
$k_i=\lambda_{i+1}-\lambda_i,\; \lambda_1=0,\; \lambda_{1+N}=\sigma.$
{}From \re{e19} a particular $q$-commutator of $A^+$ and $A^-$
may be easily fixed.

Let $A^\pm$ be well defined operators. If $\sigma>0$ then
$A^-$ is the lowering operator for
the discrete energy states. The number of normalizable independent
solutions of the equation $A^- |E\rangle=0$ determines the number ($\leq N$)
of geometric series composing the spectrum of $H$. In the $q\to 0$ limit
only few of the levels (solitons) remain in the spectrum.
If in the $q\to 1$ limit the potentials remain to be smooth, then
spectral series become equidistant.
A more detailed consideration of the $q$-transcendental potentials will
be given elsewhere [8]. Here we just mention that at $N=2$ one has
$A^-A^+-q^4A^+A^-\propto H$, which together with \re{e18} describes
a particular ``quantization" of the conformal algebra $su(1,1)$ (see, e.g.,
Ref.[9]). By the same reason this self-similar system may be called
as $q$-deformed conformal quantum mechanics.

The author is indebted to  A.Shabat for valuable discussions and helpful
comments. This research was supported by the NSERC of Canada.

\bigskip
\noindent
\medskip
{\large References}

\noindent
[1] L.C.Biedenharn, J.Phys. {\bf A22} (1989) L873;
 A.J.Macfarlane, J.Phys. {\bf A22} (1989) 4581.

\noindent
[2] A.Shabat, Inverse Prob. {\bf 8} (1992) 303.

\noindent
[3] V.Spiridonov, Phys.Rev.Lett. {\bf 69} (1992) 398.

\noindent
[4] V.Spiridonov, Mod.Phys.Lett. {\bf A7} (1992) 1241.

\noindent
[5] A.Shabat and V.Spiridonov, unpublished.

\noindent
[6] A.B.Shabat and R.I.Yamilov, Leningrad Math.J. {\bf 2} (1991) 377.

\noindent
[7] A.P.Veselov and A.B.Shabat, to be published; A.Shabat, talk at this
conference.

\noindent
[8] V.Spiridonov, in preparation.

\noindent
[9] T.L.Curtright and C.K.Zachos, Phys.Lett. {\bf B243} (1990) 237.

\end{document}